\documentclass[prl,twocolumn,showpacs,preprintnumbers,amsmath,amssymb,citeautoscript]{revtex4-1}

\usepackage{graphicx}
\usepackage{dcolumn}
\usepackage{bm}
\usepackage{color}
\usepackage{enumerate}

\newlength{\figwidth}
\begin{document}

\title{Colossal thermomagnetic response in the exotic superconductor URu$_2$Si$_2$}

\author{T.~Yamashita$^1$}
\author{Y.~Shimoyama$^{1}$}
\author{Y.~Haga$^2$}
\author{T.\,D.~Matsuda$^3$}
\author{E.~Yamamoto$^2$}
\author{Y.~Onuki$^{2,4}$}
\author{H.~Sumiyoshi$^1$}
\author{S.~Fujimoto$^5$}
\author{A.~Levchenko$^6$}
\author{T.~Shibauchi$^{1,7}$}
\author{Y.~Matsuda$^1$}

\affiliation{
$^1$Department of Physics, Kyoto University, Kyoto 606-8502, Japan\\
$^2$Advanced Science Research Center, Japan Atomic Energy Agency, Tokai 319-1195, Japan\\
$^3$Department of Physics, Tokyo Metropolitan University, Hachioji, Tokyo 192-0397, Japan\\
$^4$Department of Physics and Earth Sciences, University of the Ryukyus,  Nishihara, Okinawa 903-0213, Japan
\\
$^5$Department of Materials Engineering Science, Osaka University,  Toyonaka, Osaka 560-8531, Japan 
\\
$^6$ Department of Physics and Astronomy, Michigan State University, East Lansing, Michigan 48824, USA \\
$^7$Department of Advanced Materials Science, University of Tokyo, Kashiwa, Chiba 277-8561, Japan
}

\date{\today}


\maketitle

{\bf  
When a superconductor is heated above its critical temperature $T_c$, macroscopic coherence vanishes, leaving behind droplets of thermally fluctuating Cooper pair. This superconducting fluctuation effect above $T_c$ has been investigated for many decades and its influence on the transport, thermoelectric and thermodynamic quantities in most superconductors is well understood by the standard Gaussian fluctuation theories \cite{Larkin}. The transverse thermoelectric (Nernst) effect is particularly sensitive to the fluctuations, and the large Nernst signal found in the pseudogap regime of the underdoped high-$T_c$ cuprates\cite{Xu00,Wang06} has raised much debate on its connection to the origin of superconductivity.  Here we report on the observation of a colossal Nernst signal due to the superconducting fluctuations in the heavy-fermion superconductor URu$_2$Si$_2$.  The Nernst coefficient is enhanced by as large as one million times over the theoretically expected value within the standard framework of superconducting fluctuations.  This, for the first time in any known material, results in a sizeable thermomagnetic figure of merit approaching unity. Moreover, contrary to the conventional wisdom, the enhancement in the Nernst signal is more significant with the reduction of the impurity scattering rate.  This anomalous Nernst effect intimately reflects the highly unusual superconducting state embedded in the so-called hidden-order phase of URu$_2$Si$_2$. The results invoke possible chiral or Berry-phase fluctuations originated from the topological aspect of this superconductor, which are associated with the effective magnetic field intrinsically induced by broken time-reversal symmetry \cite{Kasa07,Yano08,Okaz10,Li13} of the superconducting order parameter.

}

The measurements of the Nernst effect provide a unique opportunity to study the superconducting fluctuations deep inside the normal state above $T_c$ \cite{Xu00,Wang06,Rull06,Pour06,Li07,Spat08,Pour09,Chan12,Dorsey,Ussi02,Ogan04,Podo07,Serb09,Mich09,Levc11,Sumi14}.  The Nernst signal $N$ is the electric field $E_y (\parallel y)$ response to transverse temperature gradient $\nabla_x T (\parallel x)$ in the presence of magnetic field $H (\parallel z)$, $N \equiv  E_y/(-\nabla_x T)$.  The Nernst coefficient defined as $\nu \equiv  N/\mu_0 H$ above $T_c$ consists of  two contributions  generated by different mechanisms: $\nu=\nu^S+\nu^N$.   The first term $\nu^S$ represents the contribution of superconducting fluctuations of either amplitude or phase of the order parameter, which is always positive.   The second term $\nu^N$ represents the contribution from the normal quasiparticles, which can be either positive or negative.  The second contribution is usually small in conventional metals. In almost all superconductors the superconducting fluctuation contribution to the  Nernst effect can be accounted for by the Gaussian-type fluctuations \cite{Dorsey,Ussi02}.    Recently,  a large Nernst signal has been reported in the pseudogap state of the underdoped high-$T_c$ cuprates, which has been discussed in terms of possible vortex-like excitations of phase disordered superconductors \cite{Xu00,Wang06,Rull06}.   Although its origin is still controversial,  these results imply that the fluctuation induced Nernst signal above $T_c$ is intimately related to the exotic superconducting state below $T_c$.  The same  conclusion is supported by the observation of an anomalous Nernst effect in the unconventional superconductor CeCoIn$_5$ \cite{Li-EPL07}. 

The heavy-fermion compound URu$_2$Si$_2$ exhibits unconventional superconductivity ($T_c\approx 1.5$\,K).  This compound is distinguished from the other heavy fermion compounds, by the fact that the mysterious hidden-order  transition takes place at $T_{\rm HO}=17.5$\,K and no evidence of magnetic order has been found below $T_{\rm HO}$ \cite{Mydo11}.  This system has been suggested to be a candidate of a chiral $d$-wave superconductor that spontaneously breaks time-reversal symmetry (TRS) in the superconducting state \cite{Kasa07,Yano08,Okaz10,Li13}.  Indeed, angular variation of the thermal conductivity and specific heat  in magnetic fields indicate the presence of point nodes in the order parameter and a chiral $d$-wave pairing symmetry in a complex form of $k_z(k_x\pm ik_y)$ has been proposed \cite{Kasa07,Yano08}.  Very recently, the broken TRS has been also reported by polar Kerr effect measurements (Kapitulnik, A. private communications).  Based on these results, possible Wyle-type topological superconducting states have been discussed \cite{Gosw13}.   It is therefore highly intriguing to examine the superconducting fluctuations in URu$_2$Si$_2$. 

\begin{figure}[t]
\includegraphics[width=1\linewidth]{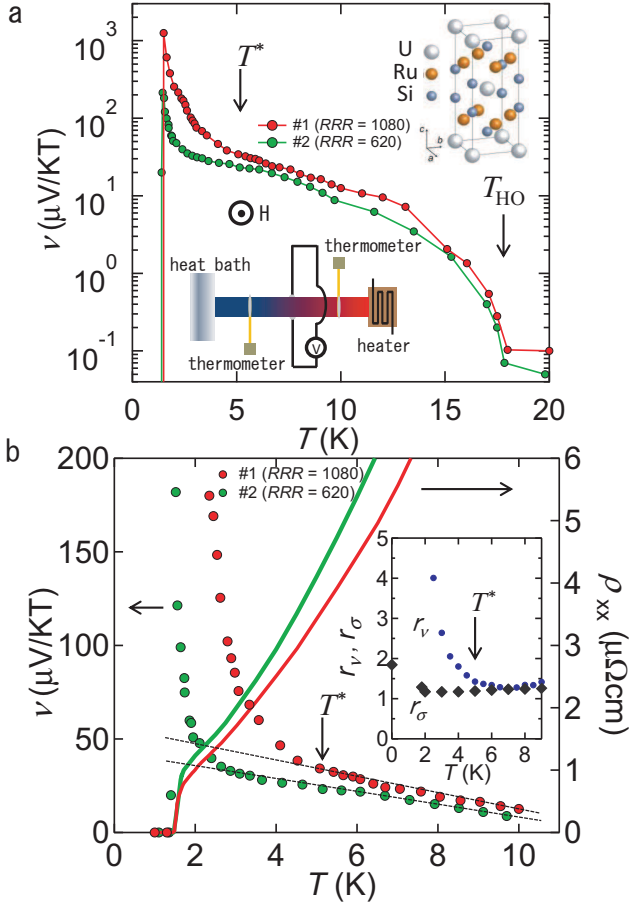}
\caption{{\bf Transverse thermoelectric response in URu$_2$Si$_2$.} {\bf a}, The $T$-dependence of Nernst coefficient $\nu$ in the zero-field limit ({\boldmath $H$}$\parallel c$) and resistivity $\rho_{xx}$  for single crystals \#1 ($RRR=1080$) and \#2 ($RRR=620$).   The $RRR$ values are determined from $\rho(300\,{\rm K})/\rho_0$ by  assuming  the $T$-dependence of $\rho_{xx}$ as $\rho_{xx}(T)=\rho_0+AT^{n}$ with $n$=1.5 and 1.7 for \#1 and \#2, respectively,  below 6\,K.    In both crystals, $T_c$ defined by the point of zero resistivity is 1.45\,K.   Upper inset illustrates the crystal structure of URu$_2$Si$_2$ and lower inset depicts the schematic measurement setup.  {\bf b},  Low temperature data of $\nu(T)$ and  $\rho_{xx}(T)$  for crystals \#1 and \#2.    Below $T^{\ast}$,  $\nu$ is largely enhanced from the $T$-linear dependence extrapolated from higher temperatures (dashed lines).   Inset shows the $T$-dependence of the ratios of Nernst coefficient and conductivity of the two crystals, $r_\nu=\frac{\nu(\#1)}{\nu(\#2)}$ and $r_{\sigma}=\frac{\rho_{xx}(\#2)}{\rho_{xx}(\#1)}$.
 } 
\end{figure}

Figure\;1a shows $\nu(T)$  in the zero-field limit  (see also Fig. S2a) and the in-plane resistivity $\rho_{xx}$ of ultraclean URu$_2$Si$_2$ single crystals \cite{Mats11}  ($T_c$=1.45\;K) with residual-resistivity-ratio ($RRR$) of 1080 (\#1) and 620 (\#2).   Above $T_{\rm HO}$, $\nu(T)$ is negligibly small and exhibits a dramatic increase on entering the hidden-order state.    Below $T^{\ast}\sim$ 5\;K,  $\nu(T)$ shows a further enhancement and  increases divergently with approaching $T_c$ (Figs.\;1a,b).  The inset of Fig.\;1b shows the $T$-dependence of the ratios of Nernst coefficient and conductivity  $\sigma=1/\rho_{xx}$ in the two crystals, $r_{\nu}\equiv \frac{\nu(\#1)}{\nu(\#2)}$ and $r_{\sigma}\equiv \frac{\rho_{xx}(\#2)}{\rho_{xx}(\#1)}$, respectively.    In contrast to  nearly $T$-independent $r_\sigma$,  $r_\nu$ increases steeply below $\sim T^{\ast}$, suggesting the appearance of an additional mechanism that generates the enhanced Nernst effect in the cleaner crystal.   These results indicate that the superconducting fluctuation effect sets in below $\sim T^{\ast}$.   As discussed later, this is supported by the $H$-dependence of the Nernst effect.   In magnetic fields,  $\nu(T)$ vanishes just below the vortex lattice melting temperature $T_{\rm melt}$ (Fig.\,2a) \cite{Okaz10}.  

We discuss the Nernst signal in the $T$-range free from the superconducting fluctuations ($T^{\ast}\alt T \alt T_{\rm HO}$).   As shown in Fig.\;2b, the increase of $RRR$ or the scattering time $\tau$ leads to an enhancement of $\nu^N$.   Within the Bolzmann theory, when $\tau$ is weakly energy dependent, $\nu^N$ can be expressed as $\nu^N=\frac{\pi^2}{3} \frac{k_B^2T}{m^{\ast}} \frac{\tau}{\varepsilon_F}$ \cite{Li07,Behn09}, where $k_B$ is the Boltzmann constant, $m^{\ast}$ is the effective mass and $\varepsilon_F$ is the Fermi energy.   

The striking enhancement of $\nu$ below $T_{\rm HO}$ is attributed to the strong reduction of $\varepsilon_F$ associated with the disappearance of carriers and concomitant enhancement of $\tau$, both of  which have been reported previously  \cite{Kasa07}.  The fact that  $r_\nu$ above $T^{\ast}$  coincides well with $r_{\sigma}$  (inset of Fig.\;1b) provides quantitative support of $\nu^N \propto \tau$. 

At lower temperatures below $T^{\ast}$, $\nu$ of clean crystals  becomes huge especially in the vicinity of $T_c$.  Indeed, $\nu$ of the cleanest crystal \#1 is comparable to that of pure semimetal Bi with  the largest Nernst coefficient reported so far \cite{Behn09}.  Moreover, the combination of the large Nernst signal and high conductivity in this system leads to a sizeable thermo-magnetic figure of merit $ZT_\epsilon=N^2\sigma T/\kappa$ ($\kappa$ is the thermal conductivity),  which quantifies the adequacy of a given material for thermoelectric refrigeration.  As shown in Fig.\,2c, this number exceeds by far the values of previously studied materials and approaches unity at 1.5\,K and 1\,T, which opens a possible route toward thermomagnetic cooling for a cryogenic Ettingshausen refrigerator \cite{Behn07}.  Interestingly, for the Nernst effect based engine there exists universal bound for the ratio between the maximum efficiency and the Carnot efficiency~\cite{Stark-PRL14}.

Now we discuss the fluctuation induced Nernst coefficient $\nu^S$.  The enhancement of $r_\nu(T)$ below $T^{\ast}$ (inset of Fig.\;1b) and  no discernible enhancement of $\nu(T)$ near $T_c$ for $RRR \sim 30$ (inset of Fig.\;2b)  indicate that $\nu^S$ is dramatically enhanced with $\tau$.    We stress that this $\tau$-dependence of $\nu^S$ is opposite to that expected in the conventional Gaussian fluctuation theories, which predict $\nu^S\propto \rho_{xx} \propto 1/\tau$  \cite{Ussi02,Serb09,Mich09,Levc11}. It has also been reported that  in underdoped cuprates the introduction of impurities by irradiation  enhances $\nu^S$ \cite{Rull06}, which is again opposite to URu$_2$Si$_2$.   Thus these results highlight an essential difference in the superconducting fluctuations between URu$_2$Si$_2$ and the other superconductors.

\begin{figure*}[t]
\includegraphics[width=0.9 \linewidth]{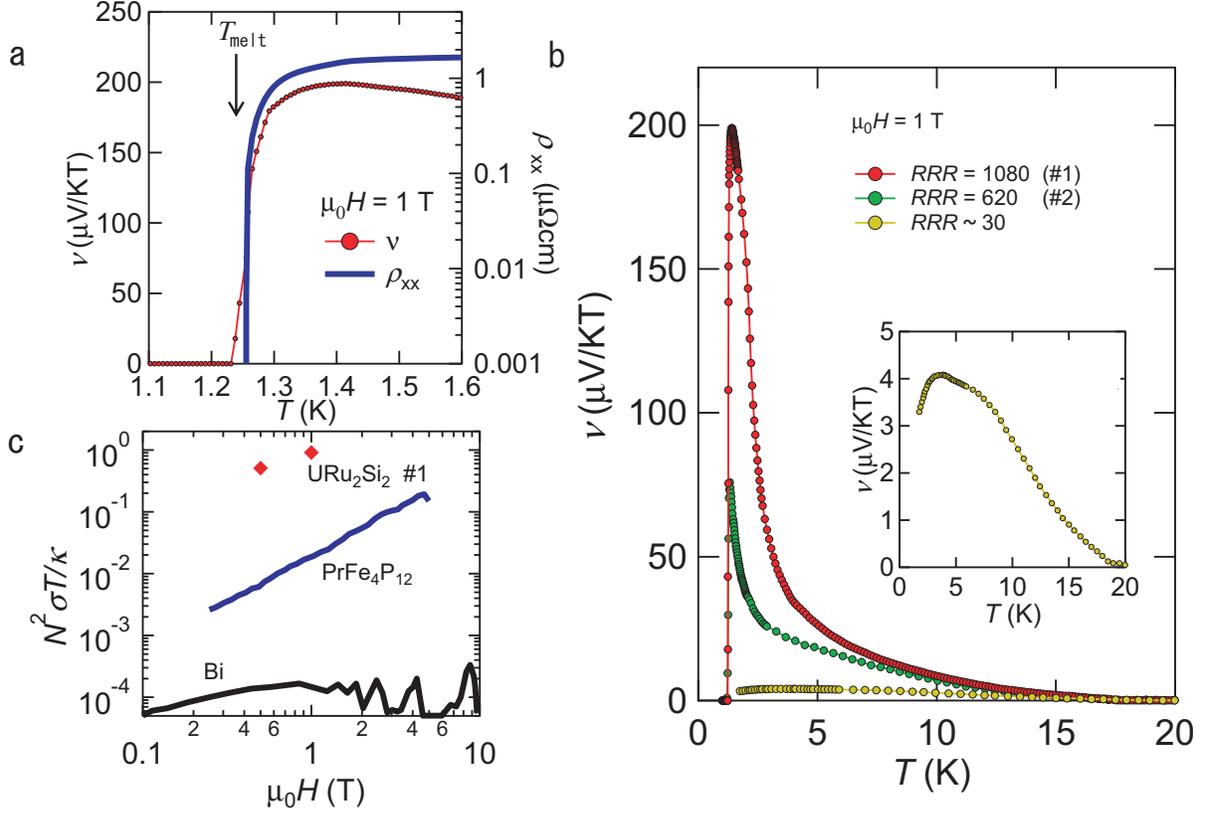}
\caption{{\bf Anomalously large Nernst signal and thermo-magnetic figure of merit.} {\bf a},  The $T$-dependence of $\nu$ (left scale) and  $\rho_{xx}$ (right scale) measured at $\mu_0H=$1\;T near the superconducting transition.  Both $\nu$ and $\rho_{xx}$ vanishes at the vortex lattice melting transition temperature $T_{\rm melt}$.     {\bf b}, Comparison of the $\nu(T)$ data at $\mu_0H=$1\;T between samples with different scattering rates ($RRR=1080$, 620 and 30). The data for $RRR \sim 30$ (expanded in the inset) is taken from Ref.\;\onlinecite{Bel04}. {\bf c}, Thermo-magnetic figure of merit $ZT_\epsilon=N^2\sigma T/\kappa$ at 1.5\,K as a function of field in crystal \#1 of URu$_2$Si$_2$ (red diamonds), which is compared with the previous data in the semimetals PrFe$_4$P$_{12}$ (blue line) and Bi (black line) at 1.2\,K taken from Ref.\;\onlinecite{Behn07}.} 
\end{figure*}

The unusual nature of the superconducting fluctuations in URu$_2$Si$_2$  is further revealed by the off-diagonal component of the thermo-electric tensor (Peltier coefficient) $\alpha_{xy}$, which is a more fundamental quantity associated with the fluctuations than the Nernst coefficient.   The relation between $\nu$ and other coefficients is given as, $\nu=\frac{1}{\mu_0 H}\left(\alpha_{xy}\rho_{xx}-S\tan\theta_H\right)$,where $\mu_0$ is the vacuum  permeability, $S$ is the Seebeck coefficient and $\tan\theta_H$ is the Hall angle.  In the whole $T$- and $H$-regions in the present study,  $\alpha_{xy}\rho_{xx} \gg S\tan\theta_H$  i.e. $\nu\approx \alpha_{xy}\rho_{xx}/\mu_0 H$ (Figs.\;S1 and S2a).    Figure\;3 shows the $T$-dependence of fluctuation-induced Peltier coefficient $\alpha_{xy}^S$ divided by $\mu_0H$ in the zero-field limit.  To determine $\alpha_{xy}^S$, the deviation from $T$-linear behaviour in $\nu(T)$ (dashed lines in Fig.\;1b) is attributed to  $\nu^S$.   The Peltier coefficient that results from the Gaussian (Aslamasov-Larkin)  fluctuations is given by,
\begin{equation}\label{alpha-xy}
\alpha_{xy}^{\rm AL}(T)=\frac{1}{12\pi}\frac{k_Be}{\hbar}\frac{\xi_{ab}^2(T)}{\ell_H^2 \xi_c(T)}\propto\frac{1}{\sqrt{\ln(T/T_c)}},
\end{equation}
where $\xi_{ab,c}(T)=\xi_{ab,c}(0)/\sqrt{\ln(T/T_c)}$ are the fluctuation coherence lengths parallel to the $ab$ plane and $c$ axis, and $\ell_H=\sqrt{\hbar /2e\mu_0 H}$ is the magnetic length.  
The blue line in Fig.\;3 shows the $T$-dependence of $\alpha_{xy}^{\rm AL}/\mu_0 H$,  calculated by using $\xi_{ab}(0)=10$\,nm and $\xi_c(0)=3.3$\,nm which are determined by the initial  slope of upper critical fields at $T_c$.  What is most spectacular is that the observed $\alpha_{xy}^S/\mu_0 H$ is  four to six orders of magnitude greater than $\alpha_{xy}^{\rm AL}/\mu_0 H$ given by Eq.\;(1).   Moreover, the $T$-dependence of $\alpha_{xy}^S(T)$  is much steeper than $\alpha_{xy}^{\rm AL}(T)$.

We stress that the observed $\alpha_{xy}^S(T)$ far exceeding $\alpha_{xy}^{\rm AL}(T)$  indeed originates from the superconducting fluctuations.  This is evidenced by the steep enhancement of both $\nu(T)$ and $r_{\nu}(T)$ below $T^{\ast}$.    The $H$-dependence of $\alpha_{xy}(H)$ provides quantitative support for this.   It has been pointed out that the size of superconducting fluctuation is set by the coherence length at low fields, while it is set  by the magnetic length at high fields. 
As a result, $\alpha_{xy}(H)$ peaks at a characteristic field $H^{\ast}$ where $\xi_{ab}(T)=\ell_H(H^{\ast})$, so that $\mu_0 H^{\ast}=\frac{\Phi_0}{2\pi \xi_{ab}^2(0)}\ln(T/T_c)$, where $\Phi_0$ is the flux quantum.  A peak field $H^{\ast}$ is called the ``ghost critical field'', and has been reported both in conventional and unconventional superconductors \cite{Pour09,Chan12}.  As shown in Fig.\;4, at all temperatures of interest, $\alpha_{xy}(H)$  exhibits a peak.  The inset of Fig.\;4 shows the peak field plotted as a function of $\ln(T/T_c)$.  The solid line, which represents $H^{\ast}(T)$ calculated by using $\xi_{ab}(0)=10$\,nm, gives a quantitative consistency with the peak field.

We discuss several possible origins for the observed colossal Nernst signal.  First, Eq.\;(1) assumes the diffusive limit,  $k_BT\ll \hbar/\tau$, while the present URu$_2$Si$_2$ appears to be in the ballistic limit,  $k_BT\gg \hbar/\tau$.  However, a ballistic theory cannot explain the observed $\alpha_{xy}^S$, which is nearly one million times greater than than $\alpha_{xy}^{\rm AL}$ (Supplementary Information).  Moreover, although such a theory shows the enhancement of $\alpha_{xy}^S$ with $\tau$, this enhancement is slower than the reduction of $\rho_{xx}$ so that $\nu^S (\propto \alpha_{xy}^S\rho_{xx}$) is still suppressed for larger $\tau$, which is inconsistent with Fig.\;2b.  Second, in the multiband system, each band with different effective coherence lengths contributes differently to the total $\alpha_{xy}^S$. To explain the observed $\alpha_{xy}^S$, however, small bands with extremely large effective coherence lengths, whose effective $H_{c2}$ corresponds to less than 1\;mOe, are required.  The multiband effect is, therefore,  highly unlikely to explain the observed $\alpha_{xy}^S$.  Third, the characteristic temperature scale of phase fluctuations is given as, $T_{\Theta}=A\frac{\hbar^2a}{4 \mu_0 k_Be^2\lambda_{ab}^2(0)}$, where $a=\sqrt{2} \xi_c(0)$, $\lambda_{ab}$ is the in-plane penetration depth, and $A$ is a dimensionless number of the order of unity \cite{Emer95}.  Using $A=2$ and $\lambda_{ab}=0.8\,\mu$m \cite{Okaz10}, we obtain $T_{\Theta}/T_c\sim100$, suggesting that the phase fluctuations are not important.

\begin{figure}[t]
\includegraphics[width=0.95\linewidth]{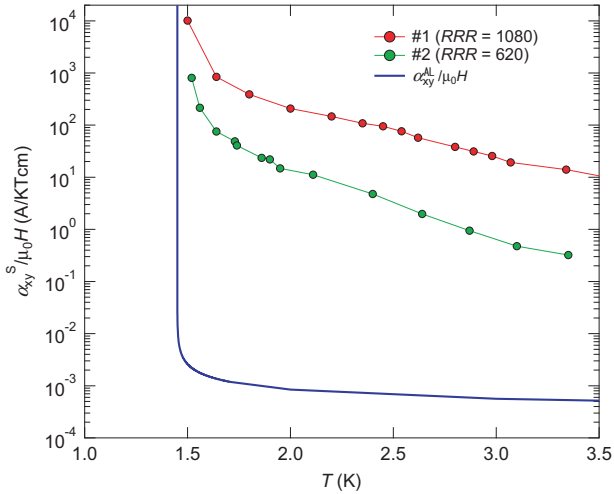}
\caption{{\bf Comparison with the standard theory of superconducting fluctuations.} The $T$-dependence of fluctuation induced Peltier coefficient  divided by magnetic field, $\alpha_{xy}^{S}/\mu_0 H$,  for crystals \#1 and \#2.   The blue line represents the Peltier coefficient that results from  Gaussian-type (Aslamasov-Larkin) fluctuations, given by Eq.\;(1).  
} 
\end{figure}

The unprecedented colossal thermomagnetic response in URu$_2$Si$_2$ appears to point to a new type of superconducting fluctuations generated by a degree of freedom which has not been hitherto taken into account. We note that it has been reported very recently that the chirality or Berry phase associated with the superconducting state with broken TRS gives rise to a new type of fluctuations \cite{Sumi14}.  In fact, according to  Ref.\;\onlinecite{Sumi14},  $\alpha_{xy}^S(T)$ is strikingly enhanced with $\tau$ and its $T$-dependence is different from that predicted by the Gaussian fluctuation theories, which are consistent with the present results at least at the qualitative level. The present results suggest that superconducting fluctuations contain a key ingredient for the topological nature of superconductors, which is a new frontier of condensed matter physics.

\begin{figure}[t]
\includegraphics[width=0.95\linewidth]{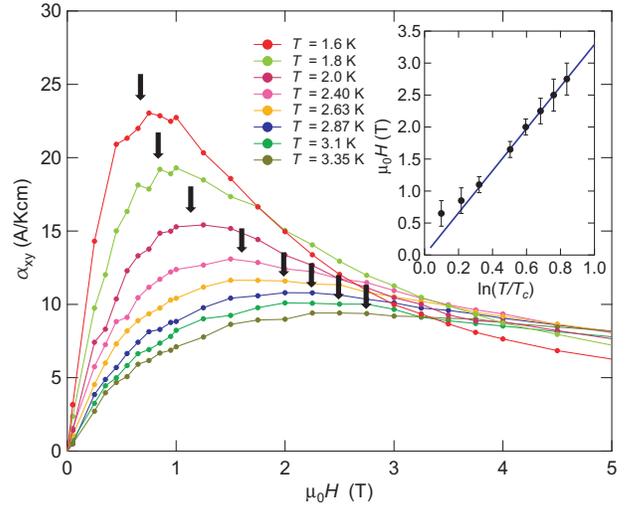}
\caption{{\bf Superconducting fluctuations and the ghost critical field.} The $H$-dependence of Peltier coefficient $\alpha_{xy}=\nu \mu_0 H/\rho_{xx}$ of \#2 crystal at several temperatures.  Arrows mark the peak fields.  The inset: Peak field plotted as a function of $\ln(T/T_c)$.  The solid line represents $\mu_0H^{\ast}=\frac{\Phi_0}{2\pi \xi_{ab}^2(0)}\ln(T/T_c)$ with $\xi_{ab}(0)=10$\;nm, which is the so-called ghost critical field.
} 
\end{figure}

\bigskip

\noindent
{\bf Methods Summary} 
The ultraclean single crystals of URu$_2$Si$_2$ were grown by the Czochralski pilling method in a tetra-arc furnace \cite{Mats11}.  The well defined superconducting transition was confirmed by the specific heat measurements.   The Nernst and Seebeck coefficients were measured by the standard dc method with one resistive heater, two Cernox thermometers and two lateral contacts (lower inset of Fig.\,1a).

\bigskip
\noindent
{\bf Supplementary Information} is available in the online version of the paper.\\

\noindent
{\bf Acknowledgements} 
We thank K. Behnia, R. Ikeda and A. Kapitulnik for fruitful discussions, and S. Tonegawa for technical assistance.  This work was supported by  Grants-in-Aid for Scientific Research (KAKENHI) from Japan Society for the Promotion of Science (JSPS), and by the ``Topological Quantum Phenomena'' (No. 25103713) Grant-in Aid for Scientific Research on Innovative Areas from the Ministry of Education, Culture, Sports, Science and Technology (MEXT) of Japan.\\

\noindent
{\bf Author contributions} 
Y.H.,  T.D.M., E.Y. and Y.O. prepared the samples. T.Y. and Y.S. carried out the measurements.  T.Y., Y.S., H.S.. S.F., A.L. T.S. and Y.M. interpreted and analysed the data.  A.L, T.S. and Y.M. wrote the manuscript.\\

\noindent
{\bf Author Information}
Reprints and permissions information is available at www.nature.com/reprints. The authors declare no competing financial interests. Correspondence and requests for materials should be addressed to Y.M.  (matsuda@scphys.kyoto-u.ac.jp).

\newpage
\begin{center}
{\large \bf Supplementary Information}
\end{center}

\renewcommand{\figurename}{Figure S$\!\!$}
\renewcommand{\tablename}{Table S$\!\!$}
\renewcommand{\theequation}{S\arabic{equation}}

\setcounter{figure}{0}
\setcounter{equation}{0}

\section{Seebeck coefficeint}

Figure\;S1 shows the longitudinal thermoelectric (Seebeck) coefficient $S$ for crystals with $RRR=1080$ (\#1), $RRR=620$ (\#2), and $RRR\sim 30$ measured at zero field. The data of $RRR\sim 30$ crystal is taken from Ref.\;26 in the main text. 
In stark contrast to Nernst coefficient shown in Fig.\;2b, the $RRR$- or $\tau$-dependence of the magnitude of $S$ is very small.  This is consistent with the results of  Boltzmann equation, in which  weak energy dependence of $\tau(\varepsilon)$ is assumed.

\begin{figure}[h]
\begin{center}
\includegraphics[width=0.8\linewidth]{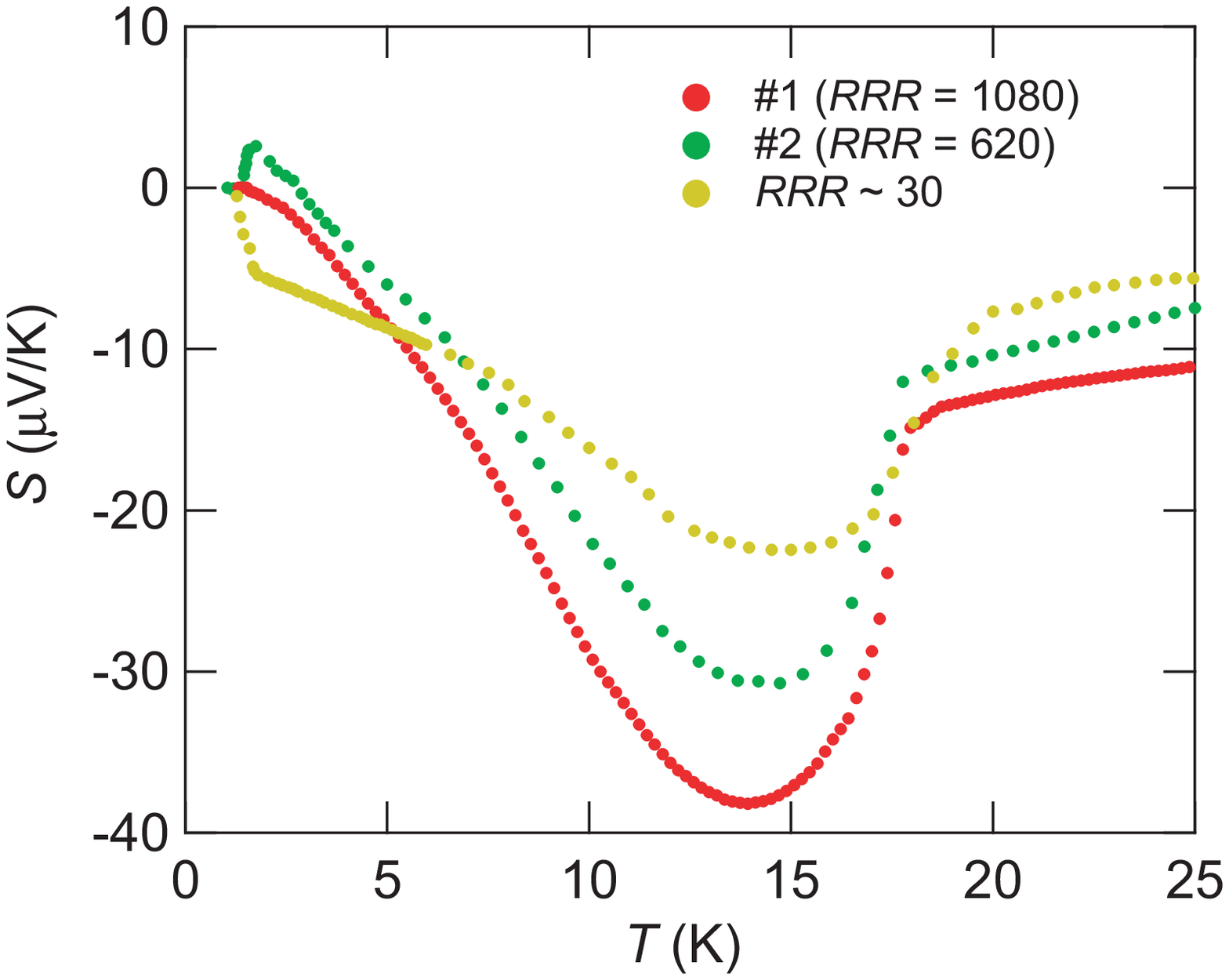}
\end{center}
\caption{Temperature dependence of the Seebeck coefficient for crystals with different $RRR$. } 
\end{figure}

\section{Field dependence of Nernst signal}

Figure\;S2a shows the $H$-dependence of the Nernst signal $N(H)$ above $T_c$ for crystal \#2.  At high temperatures, $N(H)$ increase nearly linearly with $H$.   With approaching $T_c$, $N(H)$ becomes nonlinear as a function of $H$.   We determined the Nernst coefficient $\nu\equiv N/\mu_0 H$ in the zero field limit by fitting the $H$ dependence of $N(H)$ using the polynomial functions,  and by taking a derivative $dN(H)/d(\mu_0H)$ at $H=0$.  

The Nernst signal $N=\nu \mu_0 H$ is written as,
\begin{equation}
N=\alpha_{xy}\rho_{xx}-S\tan\theta_H.
\end{equation}
Figure\;S2b depicts the $H$-dependencies of $N$ and $S\tan\theta_H$ for \#2 crystal above $T_c$.  Here $\tan \theta_H \equiv \rho_{xy}/\rho_{xx}$ is the Hall angle, where $\rho_{xx}$ and $\rho_{xy}$ are the in-plane diagonal and Hall resistivities, respectively.  Figure\;S2b shows that $\alpha_{xy}\rho_{xx}$ well dominates over $S\tan\theta_H$, indicating that $\nu$ can be approximated as $\nu\approx \alpha_{xy}\rho_{xx}/\mu_0 H$.    

\begin{figure}[t]
\begin{center}
\includegraphics[width=0.8\linewidth]{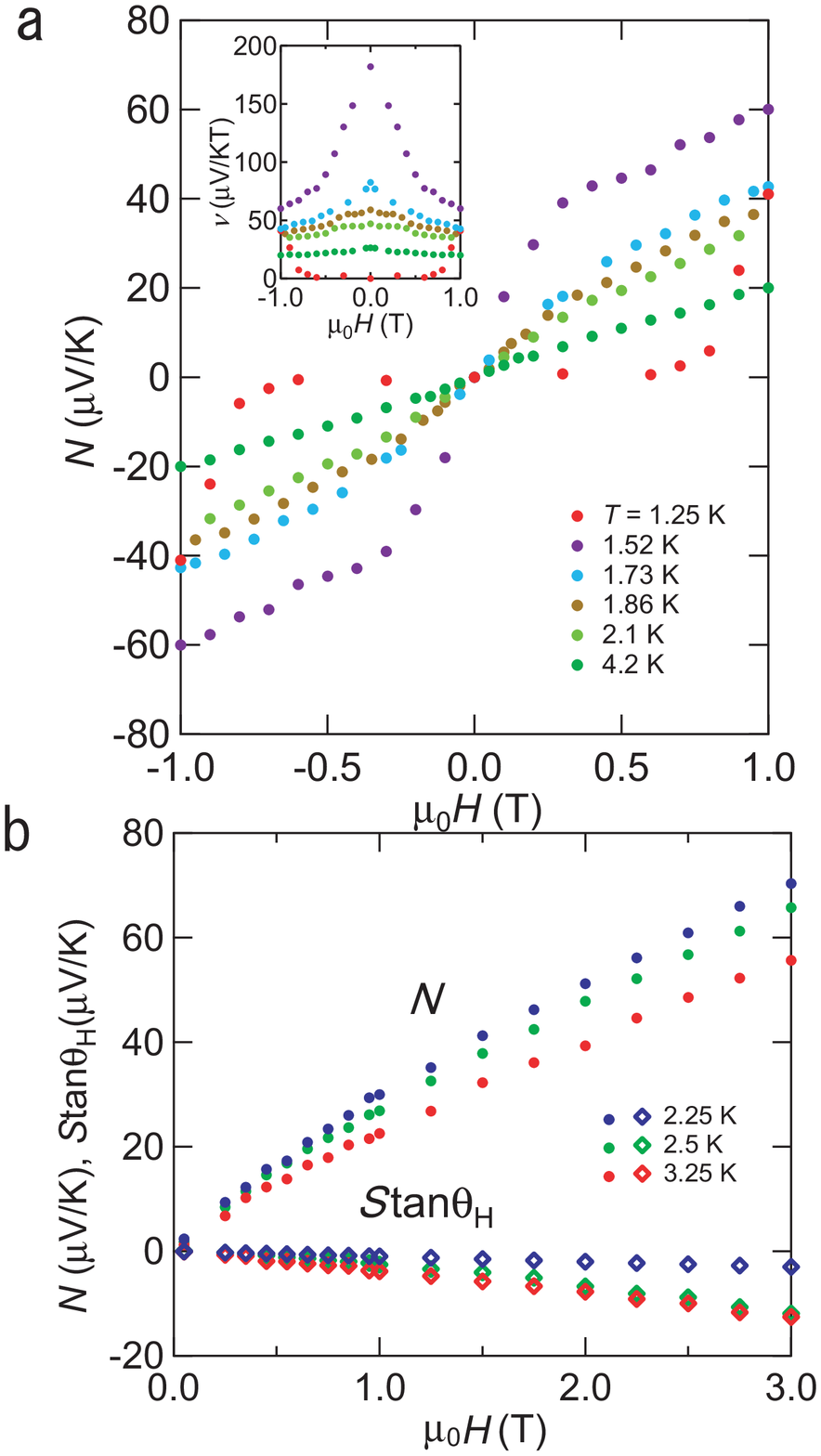}
\end{center}
\caption {{\bf a},  Field dependence of Nernst signal $N(H)$ near $T_c$ for crystal \#2.   {\bf b}, Field dependencies of the Nernst signal $N(H)$ and $S\tan \theta_H$ for crystal \#2.   } 
\end{figure}

\section{Remarks on the Peltier coefficient in the ballistic limit}

The validity of Eq.\;(1) from the main part of the paper has been microscopically established only for superconductors with the short mean free path $\ell\ll\xi(0)$, which is equivalent to the condition of diffusive scattering $T\tau\ll1$. It is important to realize that the opposite limit consists of two regimes: the ballistic limit, $\xi(0)\ll\ell\ll\xi(T)$, or equivalently $1\ll T\tau\ll\sqrt{T_c/(T-T_c)}$, and ultra-ballistic limit, $\ell\gg\xi(T)$, which is equivalent to $T\tau\gg\sqrt{T_c/(T-T_c)}$ near $T_c$. The latter case was rarely discussed in the literature despite the fact that it becomes of primary importance for the description of most experiments. It is feasible to expect that at $T\tau\gtrsim 1$ temperature dependence of the Peltier coefficient should crossover to a different law in a parameter $T\tau$, which requires separate theoretical investigation. 

We follow here the calculation of Ref.\;\onlinecite{Ussishkin-PRB03} in order to trace the $T\tau$ dependence of the transverse thermoelectric coefficient $\alpha_{xy}$. The latter is defined as
\begin{equation}\label{alpha-xy}
\alpha_{xy}=-\frac{j^Q_y}{E_xT}+\frac{cM_z}{T}=\bar{\alpha}_{xy}+\frac{cM_z}{T},
\end{equation}
where the last term $\propto M_z$ corresponds to the magnetization contribution, while the first term is the thermal response to electric field $E_x$ in the presence of the magnetic field $H$.  Thermal current $j^Q_y$  can be found through the Kubo formula (hereafter $\hbar=1$)
\begin{equation}
\frac{j^Q_y}{E_x}=-\lim_{\Omega,Q\to0}\frac{H}{c\Omega
Q}\mathrm{Re}[K_{xy}(Q,i\Omega_m)]_{i\Omega_m\to\Omega+i0},
\end{equation}
where we assumed low field limit $H\ll H_{c2}$. The response kernel is defined as follows 

\begin{eqnarray}\label{K-def}
&&K_{xy}(Q,i\Omega_m)\nonumber\\
&&=-T\sum_{q,\omega_n}\left[J^e_x(q+Q)J^e_y(q)J^Q_y(q,i\omega_n+i\Omega_m/2)L(q,i\omega_n)\right.\nonumber\\
&&~~~~~\times L(q+Q_x,i\omega_n)L(q+Q_x,i\omega_n+i\Omega_m)\nonumber\\
&&~~~~~+J^e_x(q)J^e_y(q)J^Q_y(q,i\omega_n+i\Omega_m/2)L(q,i\omega_n)\nonumber\\
&&~~~~~\times \left.L(q,i\omega_n+i\Omega_m)L(q+Q_x,i\omega_m+i\Omega_m)\right]
\end{eqnarray}

where $J^e$ and $J^Q$ are the electrical and heat current vertices, respectively, and summation goes over the Matsubara frequencies $\omega_n=2\pi nT$. Superconducting fluctuations in the clean limit are essentially nonlocal, which may strongly influence their electromagnetic response. The propagator (pair susceptibility) of the preformed Cooper pairs is of the form~\cite{Varlamov-PRB00} 

\begin{eqnarray}\label{L-def}
&&-[DL(q,i\omega)]^{-1}\nonumber\\
&&=\ln\frac{T}{T_c}+\sum^{\infty}_{n=0}\!\!
\left[\frac{1}{n+1/2}\right.\nonumber\\
&&-(\sqrt{(n+1/2+|\omega|/4\pi T+1/4\pi T\tau)^2+(v_Fq/4\pi T)^2}\nonumber\\
&&\left.-1/4\pi T\tau)^{-1}\right]
\end{eqnarray}

where $D$ is the density of states in the normal state. Calculation of $K_{xy}$ with this form of the propagator is extremely involved, however Eq.~\eqref{L-def} can be simplified by expanding it over $v_Fq/\mathrm{max}\{T,\tau^{-1}\}\ll1$ which is justified as long as $T\tau\ll\sqrt{T_c/(T-T_c)}$. Assuming this limit and generalizing Eq.~\eqref{L-def} for the anisotropic three-dimensional case one finds  
\begin{equation}\label{L}
L(q,i\omega)=-\frac{1}{D}\frac{1}{\ln\frac{T}{T_c}+\xi^2_{ab}(q^2_x+q^2_y)+\xi^2_cq^2_z+\pi |\omega|/8T}\end{equation}  
where coherence length is defined by the expression 
\begin{equation}\label{xi}
\xi^2_{ab}=\frac{\tau^2v^2_{Fab}}{3}\left[\psi\left(\frac{1}{2}\right)+\frac{1}{4\pi
T\tau}\psi'\left(\frac{1}{2}\right)-\psi\left(\frac{1}{2}+\frac{1}{4\pi
T\tau}\right)\right]
\end{equation}

\begin{figure}[t]
\begin{center}
\includegraphics[width=1\linewidth]{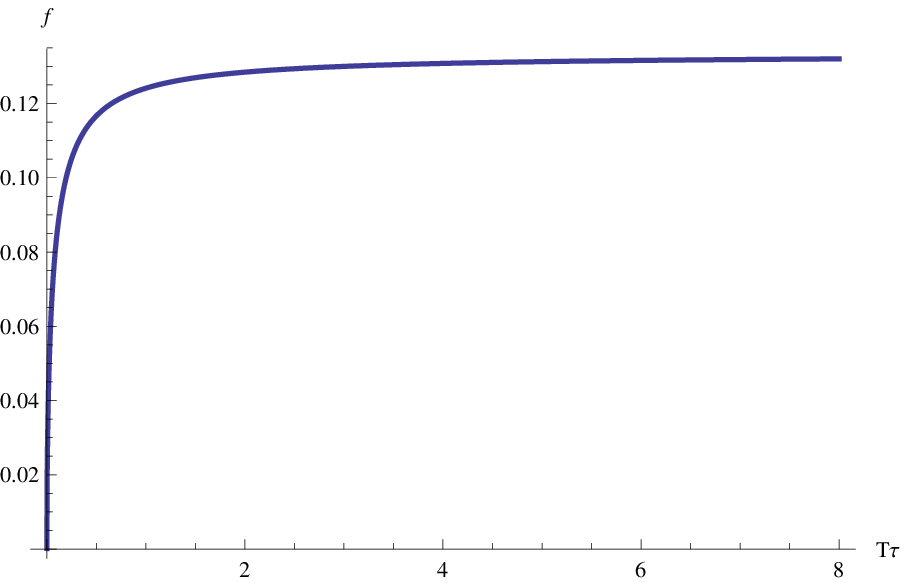}
\end{center}
\caption {Dependence of the Peltier coefficient on the scattering time $\alpha_{xy}\propto f(T\tau)$.} \label{Fig-f-T-tau}
\end{figure}

with $\psi(z)$ being the Euler digamma function. The coherence length $\xi_c$ along the $c$-axis has the same functional form but different Fermi velocity $v_{Fc}$ (for simplicity we took scattering rate to be independent of the direction). In this ballistic limit both current and heat vertices are still local 
\begin{eqnarray}
&&J^e_i(q,i\omega)=2eB_i(q,i\omega),\quad J^Q_i(q,i\omega)=-i\omega B_i(q,i\omega), \nonumber\\
&&\hspace{20mm}B_{x,y}(q,i\omega)=-2N\xi^2_{ab}q_{x,y}.
\end{eqnarray}
$B_z$ is obtained by replacing $\xi_{ab}\to\xi_c$. Having these expressions we perform summation over the Matsubara frequency in Eq.~\eqref{K-def} followed by an analytical continuation which gives for the thermal part of the Peltier coefficient 
\begin{eqnarray}
&&\bar{\alpha}_{xy}=\frac{4e^2H}{\pi
cT}\sum_qB^2_x(q)B^2_y(q)\!\!\int\!\! d\omega\coth\frac{\omega}{2T}\left[(\Re L^R(q,\omega))^3\right. \nonumber\\
&&\hspace{10mm}\times \left.\Im L^R(q,\omega)+\Re L^R(q,\omega)(\Im
L^R(q,\omega))^3\right]
\end{eqnarray}
where $L^R$ is the retarded component of Eq.~\eqref{L}. In the immediate vicinity of superconducting transition $(\xi q)^2\sim\omega/T\sim\ln(T/T_c)\ll1$ so that one can safely
approximate $\coth(\omega/2T)\approx2T/\omega$. Next, it is convenient to rescale all the momenta in units of coherence length $\xi_{ab}q_x=\kappa_x, \xi_{ab}q_y=\kappa_y,
\xi_cq_z=\kappa_z$ and pass to the spherical coordinates in the new
momentum variable $\kappa$, namely
$\kappa_x=\kappa\sin\theta\sin\phi,
\kappa_y=\kappa\sin\theta\cos\phi, \kappa_z=\kappa\cos\theta$,
which gives us
\begin{eqnarray}
&&\bar{\alpha}_{xy}=\frac{8e}{\pi\ell^2_H}\frac{1}{\xi^2_{ab}\xi_c}\nonumber\\
&&\hspace{10mm}\int\frac{\kappa^2\sin\theta
d\kappa d\theta
d\phi}{8\pi^3}16D^4\xi^4_{ab}\kappa^4\sin^4\theta\sin^2\phi\cos^2\phi\nonumber\\
&&\hspace{10mm}\int\frac{d\omega}{\omega}\left[(\Re L^R)^3\Im L^R+\Re L^R(\Im
L^R)^3\right]
\end{eqnarray}
where magnetic length is $\ell_{H}=\sqrt{c/eH}$. 
As a final step we introduce dimensionless variables  
$\lambda=\ln(T/T_c)$, $x=\kappa^2$, $y=\pi\omega/8T$, which implies for the propagator 
\begin{equation}
\Re L^R=-\frac{1}{D}\frac{\lambda+x}{(\lambda+x)^2+y^2},\qquad \Im
L^R=-\frac{1}{D}\frac{y}{(\lambda+x)^2+y^2}.
\end{equation}
and use integrals 
\begin{eqnarray}\nonumber
&&\int^{2\pi}_{0}\cos^2\phi\sin^2\phi d\phi=\frac{\pi}{4},\quad
\int^{\pi}_{0}\sin^5\theta d\theta=\frac{16}{15},\quad \nonumber\\
&&\int^{+\infty}_{-\infty}\frac{dy}{(1+y^2)^3}=\frac{3\pi}{8},\quad
\int^{\infty}_{0}\frac{x^{5/2}dx}{(1+x)^4}=\frac{5\pi}{16}\nonumber
\end{eqnarray}
to obtain Peltier coefficient for the range of scattering satisfying $1\ll T\tau\ll\sqrt{T_c/(T-T_c)}$
\begin{equation}
\bar{\alpha}_{xy}=\frac{k_Be}{4\pi\hbar}\frac{\epsilon\xi_{ab}(0)}{\ell^2_H}\frac{T_c}{T}\frac{f(T\tau)}{\sqrt{\ln(T/T_c)}}
\end{equation}
here $\epsilon=\xi_{ab}/\xi_c$ is the anisotropy parameter and dimensionless function is 
\begin{equation}
f(z)=\frac{z}{\sqrt{3}}\sqrt{\psi\left(\frac{1}{2}\right)+\frac{1}{4\pi
z}\psi'\left(\frac{1}{2}\right)-\psi\left(\frac{1}{2}+\frac{1}{4\pi
z}\right)}.
\end{equation}
Magnetization contribution $cM_z/T$ is of the same form as $\bar{\alpha}_{xy}$ but opposite in sign and comes with the coefficient $1/6\pi$ so that an overall coefficient in $\alpha_{xy}$ of Eq.\;\eqref{alpha-xy} is $1/12\pi$. Function $f(T\tau)$ governs dependence of $\alpha_{xy}$ on a scattering time which displays sharp growth by a factor of six followed by a rapid saturation at $T\tau\gtrsim1$, see Fig.\;S\ref{Fig-f-T-tau} for the illustration.

\end{document}